# Engineering quantum spin Hall insulators by strained-layer heterostructures

T. Akiho[†], F. Couëdo[†], H. Irie, K. Suzuki, K. Onomitsu, and K. Muraki[*]
NTT Basic Research Laboratories, NTT Corporation, 3-1 Morinosato-Wakamiya, Atsugi 243-0198, Japan

Quantum spin Hall insulators (QSHIs), also known as two-dimensional topological insulators, have emerged as an unconventional class of quantum states with insulating bulk and conducting edges originating from nontrivial inverted band structures, and have been proposed as a platform for exploring spintronics applications and exotic quasiparticles related to the spin-helical edge modes. Despite theoretical proposals for various materials, however, experimental demonstrations of QSHIs have so far been limited to two systems—HgTe/CdTe and InAs/GaSb—both of which are lattice-matched semiconductor heterostructures. Here we report transport measurements in yet another realization of a band-inverted heterostructure as a QSHI candidate—InAs/In$_x$Ga$_{1-x}$Sb with lattice mismatch. We show that the compressive strain in the In$_x$Ga$_{1-x}$Sb layer enhances the band overlap and energy gap. Consequently, high bulk resistivity, two orders of magnitude higher than for InAs/GaSb, is obtained deep in the band-inverted regime. The strain also enhances bulk Rashba spin-orbit splitting, leading to an unusual situation where the Fermi level crosses only one spin branch for electronlike and holelike bands over a wide density range. These properties make this system a promising platform for robust QSHIs with unique spin properties and demonstrate strain to be an important ingredient for tuning spin-orbit interaction.

[†]These authors contributed equally to this work.

[*]corresponding author: muraki.koji@lab.ntt.co.jp

Quantum spin Hall insulators (QSHIs) have insulating bulk and conducting edges that originate from topologically nontrivial inverted band structures and are protected from backscattering by time-reversal symmetry[1-8]. Owing to these unique properties of the spin-helical edge modes, QSHIs have been proposed as a platform for exploring spintronics applications[1-3, 8] and exotic quasiparticles useful for topological quantum computation[9, 10]. Despite theoretical predictions for various materials[4-6, 11-19], however, experimental demonstrations of QSHIs have so far been limited to two systems—HgTe/CdTe[6-8] and InAs/GaSb[12, 20-23]—both of which are lattice-matched semiconductor heterostructures. InAs/GaSb quantum wells (QWs), characterized by a broken-gap type-II band alignment, have recently been attracting increasing interest fueled mainly by their favorable properties, including their good interface with superconductors[24, 25] and in-situ electric tunability[26, 27]. However, the residual bulk conductivity associated with the small energy gap in this system has been an obstacle to unambiguous identification of edge properties even at low temperatures[28]. In this Letter, we propose a strained InAs/In$_x$Ga$_{1-x}$Sb QW structure as a QSHI candidate and demonstrate that the strain enhances the energy gap and leads to superior bulk insulation properties.

Figure 1(a) shows the band-edge profile of InAs/In$_x$Ga$_{1-x}$Sb ($x$ = 0.25) QWs pseudomorphically grown on AlSb. As in the conventional InAs/GaSb system, electrons and holes are separately confined to the InAs and InGaSb wells, respectively, as a result of the staggered band-gap alignment. The conduction band bottom of InAs is located ~0.17 eV below the valence band top of In$_{0.25}$Ga$_{0.75}$Sb. Consequently, when the thickness of the InAs and InGaSb layers ($t_{InAs}$ and $t_{InGaSb}$, respectively) are such that quantum confinement is not too strong, the system is in the band-inverted regime, with the first electron subband E1 located below the first heavy-hole subband HH1.

In$_{0.25}$Ga$_{0.75}$Sb has a 0.82% lattice mismatch with respect to AlSb, which induces compressive strain in the InGaSb layer. The shear strain component enhances the heavy-hole (HH) light-hole (LH) splitting at the Γ point [Fig. 1(b)], moving the HH (LH) band upwards (downwards) and also strengthening the Γ$_8$–Γ$_6$ band inversion[29]. As we show below, these impact the energy dispersion in the topological phase. Figures 1(c)–(f) compare the band structure of strained InAs/In$_{0.25}$Ga$_{0.75}$Sb QWs [Figs. 1(d), (f)] with that of unstrained InAs/GaSb QWs [Figs. 1(c), (e)], obtained from the 8-band k·p calculations for $t_{InAs}$ = 10 nm and $t_{(In)GaSb}$ = 6 nm with strain effects taken into account[30] (material parameters are from Ref. [31]). Hybridization of E1 and HH1 subbands away from the Γ point results in anticrossing at a finite $k$ (hereafter denoted as $k_{cross}$)[12]. The QSHI phase emerges when the Fermi level $E_F$ is tuned to this hybridization gap Δ. As the calculations show, the stronger Γ$_8$–Γ$_6$ band inversion results in a larger band overlap $E_{g0} \equiv E_{HH1} - E_{E1}$ and, accordingly, a larger $k_{cross}$ in the strained QWs. At the same time, Δ is enhanced from 4.5 to 10.8 meV. As Figs. 1(e) and (f) show, the latter is due to the strain-enhanced HH-LH splitting, which works to prevent unwanted LH-HH level interaction that reduces Δ. Furthermore, the increased $E_{g0}$ implies that the QSHI phase can be realized for smaller $t_{InAs}$ and $t_{InGaSb}$ and hence for stronger interlayer coupling, allowing Δ to be further enhanced. Interestingly, strain also enhances the Rashba spin splitting[32] near the anticrossing, both in the conduction and valence bands.

The heterostructures studied were grown by molecular-beam epitaxy on Si-doped (001) GaAs substrates. The QWs comprise InAs (top) with thickness $t_{InAs}$ and 5.9-nm-thick $In_{0.25}Ga_{0.75}Sb$ (bottom), sandwiched between an 800-nm-thick AlSb buffer layer and a 50-nm-thick AlSb upper barrier. We first present data for $InAs/In_{0.25}Ga_{0.75}Sb$ QW with $t_{InAs}$ = 10.9 nm, measured in a Hall bar geometry (width $W$ = 50 μm and voltage-probe distance $L$ = 180 μm) with front and back gates. (The back-gate voltage $V_{BG}$ was kept at 0 V throughout this work. See supplementary material for details about the sample.) The longitudinal resistance $R_{xx}$ measured at $B$ = 0 T and $T$ = 2 K as a function of front-gate voltage $V_{FG}$ exhibits a peak at $V_{FG}$ = –0.18 V [Fig. 2(a), lower panel]. With a finite magnetic field $B$ (= 1.5 T) applied perpendicular to the sample, the sign of the Hall resistance $R_{xy}$ changes as $V_{FG}$ is swept across the $R_{xx}$ peak [Fig. 2(a), upper panel], indicating a change in the majority carrier type from holes to electrons.

We determined the density of electrons and holes separately through magnetotransport measurements. In Fig. 2(b), we show $R_{xx}$ vs $B$ traces for several values of $V_{FG}$. Fast Fourier transform (FFT) analysis of the Shubnikov-de Haas (SdH) oscillations reveals two $1/B$ frequency components $f_{1/B}$ as shown in the insets. The one with the corresponding carrier density increasing (decreasing) with $V_{FG}$ can be identified as being associated with electrons in InAs (holes in InGaSb) [Fig. 2(c), lower panel].

We note that not only the electron density $n_e$ but also the hole density $n_h$ changes with $V_{FG}$. This arises from the finite density of states $D_{InAs}$, or quantum capacitance $c_{InAs} = e^2 D_{InAs}$, of the electron subband in the InAs well, where $D_{InAs} = (g_s/2) m_e^*/\pi\hbar^2$ with $m_e^*$ the electron effective mass and $g_s$ the spin degeneracy ($e$ is the elementary charge and $\hbar = h/2\pi$ is the reduced Planck constant). That is, due to the finite $D_{InAs}$, a portion of electrons added to the InAs well upon increasing $V_{FG}$ is transferred to the InGaSb well, which decreases $n_h$. The equivalent circuit model[26] shown in the upper inset of Fig. 2(c) accounts for the variations of $n_e$ and $n_h$ with $V_{FG}$ and thus provides a good fit with $g_s(m_e^*/m_0) = 0.052$ as shown by the solid lines (see supplementary material for other parameters). A linear fitting of $n_e$ and $n_h$ allows us to locate the charge neutrality point (CNP), where $n_e = n_h$, at $V_{FG}$ = –0.17 ± 0.01 V. This confirms that the $R_{xx}$ peak is located at the CNP. [The finite $R_{xy}$ at the CNP in Fig. 2(a) is due to the admixture of $R_{xx}$ into $R_{xy}$, which becomes discernable since $|R_{xy}| \ll R_{xx}$ near the CNP.]

Carrier density $n$ is related to $f_{1/B}$ as $n = g_s(e/h)f_{1/B}$. A striking observation revealed by the magnetotransport data in Fig. 2(b) is that both the electron and hole subbands are fully spin split (i.e., $g_s$ = 1). In the upper panel of Fig. 2(c), we compare the net charge carrier density $n_{net} = n_e - n_h$ obtained from the FFT analysis of the SdH oscillations with the carrier density $n_{Hall} = B/eR_{xy}$ deduced from $R_{xy}$ at $B$ = 5.0 T. If we assume $g_s$ = 2, the resultant $n_{net}$ turns out to be twice as large as $n_{Hall}$. As Fig. 2(c) shows, $n_e$ and $n_h$ deduced with $g_s$ = 1 instead provide $n_{net}$ consistent with $n_{Hall}$, which demonstrates spin-split Landau levels. We ascribe the complete spin splitting of the Landau levels to the large Rashba spin splitting (at $B = 0$ T) that arises from the structural inversion asymmetry[32] inherent to the InAs/(In)GaSb system. When $E_F$ is located at the position shown in the lower inset of Fig. 2(c), for a given **k** direction $E_F$ intersects only one spin branch for electronlike and holelike bands. This accounts for the disappearance of the spin degree of freedom in the SdH oscillations. (We add that $g_s = 1$ applies only near the CNP. The behavior at higher $V_{FG}$, where the upper spin branch is occupied, will be reported elsewhere.) With $g_s = 1$, we obtain $n_e$ and $n_h$ at the CNP (denoted as $n_{cross}$) to be $3.6 \times 10^{15}$ m$^{-2}$, where $n_{cross}$ is related to $k_{cross}$ as $k_{cross} = (4\pi n_{cross}/g_s)^{1/2}$ and is a measure of the band overlap. Noting that $g_s = 1$ in our case, we obtain $k_{cross} = 0.21$ nm$^{-1}$, which would correspond to a very large $n_{cross}$ of $7.2 \times 10^{15}$ m$^{-2}$ in spin-degenerate systems. It is to be emphasized that we determined $n_{cross}$ using a linear fit in the two-carrier regime [Fig. 2(c)] (i.e., not a linear extrapolation from the single-carrier regime). This is essential for accurate evaluation of the band overlap.

We performed similar measurements and analysis for InAs/InGaSb QWs with varying $t_{InAs}$ (= 8.5, 9.1, and 10.0 nm) and constant $t_{InGaSb}$ of 5.9 nm. The values of $n_{cross}$ obtained for these samples are plotted in Fig. 3 as a function of $t_{InAs}$ and compared with calculation. For comparison with the spin-degenerate systems, here we take the vertical axis to be $n_{cross}/g_s$ (i.e., the CNP carrier density per spin species). The calculation shows that $n_{cross}$ sensitively depends on the lattice constant of the buffer layer. Our data fall within the range of 94–98% strain relaxation of the AlSb buffer layer with respect to the GaAs substrate (shown in green). The calculation assuming the same strain relaxation yields much lower $n_{cross}/g_s$ for the InAs/GaSb system (shown in blue). This is consistent with our control experiment that indicated low $n_{cross}$ below our resolution for InAs/GaSb QWs with $t_{InAs} = 10$ nm. These results clearly demonstrate the enhanced band overlap in the InAs/InGaSb strained QWs.

Now we turn our attention to the resistance peak height at the CNP, which mainly reflects in-gap states in the bulk. Measurements at $T = 0.25$ K show that the $R_{xx}$ peak grows with decreasing $t_{InAs}$, reaching $R_{peak} = 889$ kΩ for $t_{InAs} = 8.5$ nm [Fig. 4(a)]. When the conduction is dominated by the bulk, the resistivity can be estimated as $\rho_{xx} = (W/L)R_{peak}$. Using a Corbino geometry or a special sample geometry in which edge conduction can be neglected[23], we confirm this assumption to be valid except for the narrowest QW, for which we obtain higher $\rho_{xx}$ of $15.7h/e^2$ as compared to $(W/L)R_{peak} \sim 10h/e^2$ estimated from the Hall bar device. This suggests that edge conduction is not negligible for the $t_{InAs} = 8.5$ nm QWs even in a macroscopic Hall bar device, which reflects the high bulk resistivity.

In Fig. 4(b), we plot $\rho_{xx}$ at the CNP of the InAs/InGaSb QWs as a function of $(2/g_s)n_{cross}$ together with the data reported for InAs/GaSb QWs[26, 28, 33-37]. Here we take the horizontal axis to be $(2/g_s)n_{cross}$ so that data for the same $k_{cross}$ can be compared. Quantum transport theory predicts that for $\Gamma_\phi \ll \Delta \ll E_{g0}$ the bulk resistivity at $T = 0$ K scales as $\rho_{xx} \sim (\Delta/E_{g0})h/e^2$, where $\Gamma_\phi$ is the level broadening[38]. As shown by the black solid (dashed) lines, $\Delta/E_{g0}$ evaluated with the 8-band **k·p** calculation (constant $\Delta$ of 5 meV) provides an upper bound of $\rho_{xx}$ reported for InAs/GaSb QWs in the large-$k_{cross}$ regime[28]. In contrast, $\rho_{xx}$ values measured for the InAs/InGaSb QWs significantly exceed $(\Delta/E_{g0})h/e^2$ evaluated for this system (red solid line). They are two orders of magnitude higher than those for the InAs/GaSb QWs with the same $k_{cross}$, being comparable to the high values in the small-$k_{cross}$ regime obtained by intentionally introducing disorder[36].

The mechanism that dictates $\rho_{xx}$ for InAs/InGaSb QWs at low temperatures is not known at present. Disorder is unlikely to be the dominant factor, as our InAs/InGaSb QWs have electron mobility (~ 5 m$^2$/Vs at $n_e = 1.5 \times 10^{16}$ m$^{-2}$) comparable to the values reported for InAs/GaSb systems. The temperature dependence of $\rho_{xx}$ suggests log$T$ behavior (not shown), rather than thermal activation. Nevertheless, it is useful to compare the behavior of $\rho_{xx}$ with the calculated $t_{InAs}$ dependence of $\Delta$, which indicates that $\Delta$ increases with decreasing $t_{InAs}$ from 10.9 to 8.5 nm (Fig. 5). This suggests that the bulk resistivity can be increased further by engineering the heterostructure in such a way that $\Delta$ is maximized. The calculation predicts that $\Delta$ as large as 25 meV (~ 290 K) can be achieved for a highly strained QW with $x = 0.40$ pseudomorphically grown on GaSb (2.48% strain).

Our results demonstrate that an InAs/InGaSb strained QW structure is a promising platform for robust QSHIs, where intrinsic edge physics can be studied with the advantage of good bulk insulation deep in the band-inverted regime without doping to increase disorder. The density range over which complete spin polarization is observed suggests a spin splitting greater than ~ 15 meV, which by far exceeds that calculated for the conduction band (~ 1 meV). The huge Rashba splitting warrants further study to clarify the role of strain in tuning spin-orbit interaction and explore unconventional transport expected in this regime[39].

See supplementary material for details about the sample and the parameters used in the fitting.

The authors thank H. Murofushi for his help during the sample processing. This work was supported by JSPS KAKENHI Grant Numbers JP15H05854, JP26287068.

Figure Captions

FIG. 1 (a) Band edge profile of InAs/In$_{0.25}$Ga$_{0.75}$Sb quantum wells (QWs) with AlSb barriers, assuming pseudomorphic growth on AlSb. (b) Close-up of the band alignment at the InAs/In$_{0.25}$Ga$_{0.75}$Sb interface. The dotted lines represent the position of the energy levels in the absence of strain. The valence band offset between InGaSb and InAs was defined for the "center of mass" of the HH and LH band (i.e., without the shear strain term) and was taken to be 0.56 eV, equal to that between GaSb and InAs. (c)–(f) In-plane energy dispersions of unstrained InAs/GaSb QWs (c, e) and strained InAs/In$_{0.25}$Ga$_{0.75}$Sb QWs (d, f), obtained from 8-band **k·p** calculations. (c) [(d)] shows a close-up of the region marked by the green rectangle in (e) [(f)]. The shaded regions in (c) and (d) represent the energy gap for each **k** direction. To take into account the band anisotropy between the [100] and [110] directions, we define the hybridization gap Δ as the energy difference between the lower of the upper band minima and the higher of the lower band maxima.

FIG. 2 Transport properties of InAs/In$_{0.25}$Ga$_{0.75}$Sb quantum wells with $t_{InAs}$ = 10.9 nm. (a) Longitudinal resistance $R_{xx}$ at $B$ = 0 T (lower panel) and Hall resistance $R_{xy}$ at $B$ = 1.5 T (upper panel) measured at $T$ = 2 K. (b) $R_{xx}$ vs $B$ traces for various values of front-gate voltage $V_{FG}$. The insets show fast Fourier transform (FFT) amplitude of the Shubnikov de-Haas (SdH) oscillations at $V_{FG}$ = 0.5 and −1.0 V. (c) (lower panel) $V_{FG}$ dependence of the two 1/$B$ frequency components (right axis) and the corresponding carrier density calculated assuming spin-split Landau levels (left axis). The solid lines are fits to the data obtained using the equivalent circuit model shown in the upper inset. The lower inset illustrates the situation where the Fermi level intersects only one spin branch for electronlike and holelike bands for a given **k** direction. (c) (upper panel) $V_{FG}$ dependence of net charge carrier density $n_{net}$ = $n_e$ − $n_h$ obtained from the FFT analysis of the SdH oscillations and its comparison with the carrier density $n_{Hall}$ deduced from $R_{xy}$ at $B$ = 5 T. The black solid line is a linear fit to $n_{net}$ vs $V_{FG}$.

FIG. 3 Carrier density at the charge neutrality point ($n_{cross}$) for InAs/In$_{0.25}$Ga$_{0.75}$Sb quantum wells with different InAs layer thickness $t_{InAs}$ The vertical axis is taken to be $n_{cross}/g_s$, where $g_s$ is the spin degeneracy. The solid line represents the value expected from **k·p** calculation assuming pseudomorphic growth on AlSb (i.e., 100% strain relaxation of the AlSb buffer layer). The dashed line shows the same calculation, but for pseudomorphic growth on GaSb. The green shaded region demarcates the range of 94–98% strain relaxation of the AlSb buffer layer with respect to the GaAs substrate. The blue shaded region delineates the range of corresponding $n_{cross}/g_s$ values for the InAs/GaSb system assuming the same strain relaxation in the buffer layer.

FIG. 4 (a) $R_{xx}$ vs $V_{FG}$ at $T = 0.25$ K for InAs/In$_{0.25}$Ga$_{0.75}$Sb QWs with different $t_{InAs}$. (b) Summary of $\rho_{xx}$ values at the charge neutrality point measured for InAs/In$_{0.25}$Ga$_{0.75}$Sb QWs and those reported for InAs/GaSb QWs, plotted as a function of $(2/g_s)n_{cross}$. The horizontal axis is taken to be $(2/g_s)n_{cross}$ ($g_s$ is the spin degeneracy) so that data for the same $k_{cross}$ (top axis) can be compared. Open symbols represent values deduced from measurements on Hall bar devices, whereas closed symbols are those obtained from Corbino devices or in a geometry in which edge conduction can be neglected. The solid lines represent $\Delta/E_{g0}$ evaluated from **k·p** calculations for unstrained InAs/GaSb (black) and InAs/In$_{0.25}$Ga$_{0.75}$Sb (red) QWs pseudomorphic on AlSb, both with $t_{(In)GaSb} = 6$ nm. The black dashed line shows $\Delta/E_{g0}$ evaluated with a constant $\Delta$ of 5 meV and $E_{g0}$ calculated as $E_{g0} = (1/2m_e^* + 1/2m_h^*)\hbar^2 k_{cross}^2$ by using the electron and hole effective masses $m_e^* = 0.04m_0$ and $m_h^* = 0.09m_0$.

FIG. 5 Hybridization gap $\Delta$ of InAs/In$_x$Ga$_{1-x}$Sb QWs ($x = 0$, 0.25, and 0.40) calculated as a function of InAs layer thickness $t_{InAs}$. In$_x$Ga$_{1-x}$Sb layer thickness is 6 nm. The blue solid line shows the calculation for unstrained QW with $x = 0$. Red and black solid (dashed) lines show calculations assuming pseudomorphic growth on AlSb (GaSb). The inset shows the energy dispersion for a QW with $x = 0.40$, $t_{InAs} = 6.5$ nm, and $t_{InGaSb} = 6$ nm, pseudomorphic on GaSb, where $\Delta$ reaches 25 meV (~ 290 K).

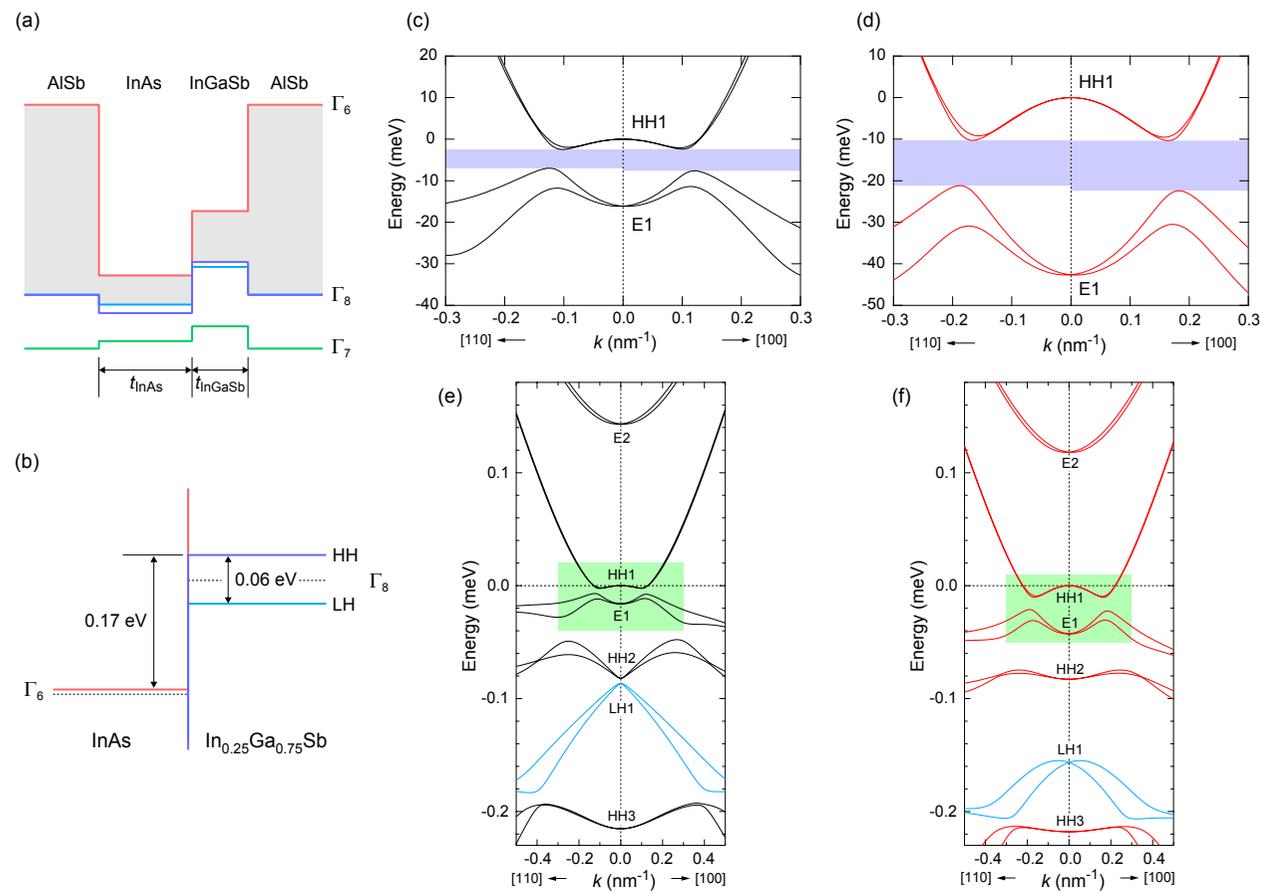

Fig. 1

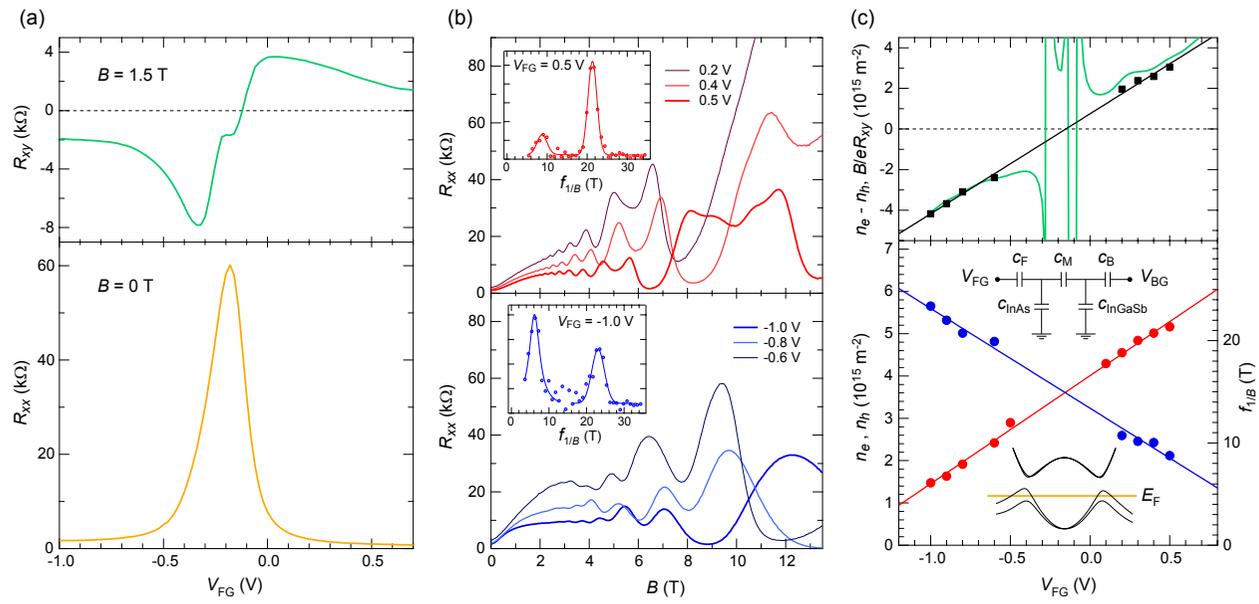

Fig. 2

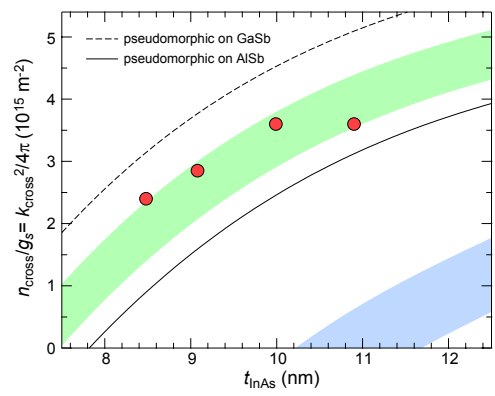

Fig.3

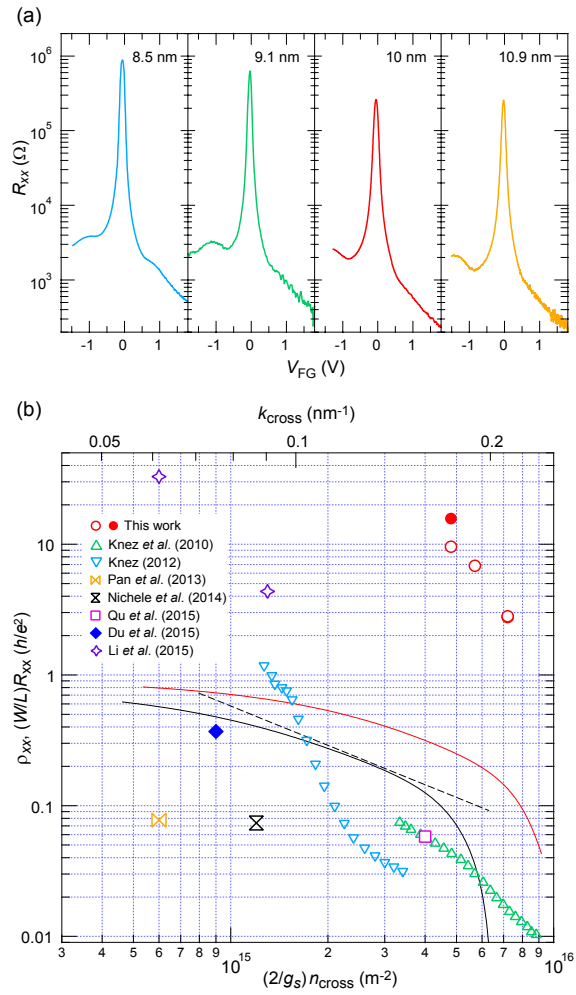

Fig. 4

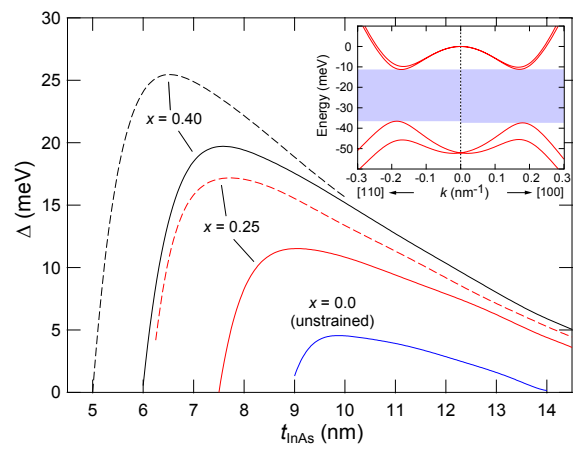

Fig. 5

Supplementary Material for

# Engineering quantum spin Hall insulators by strained-layer heterostructures

T. Akiho, F. Couëdo, H. Irie, K. Suzuki, K. Onomitsu, and K. Muraki

NTT Basic Research Laboratories, NTT Corporation, 3-1 Morinosato-Wakamiya, Atsugi 243-0198, Japan

Sample fabrication

The heterostructures were grown by molecular-beam epitaxy on Si-doped (001) GaAs substrates. The layer structure (from the bottom to the surface) comprises a 600-nm-thick Si-doped ([Si] = $10^{18}$ cm$^{-3}$) GaAs buffer layer, 800-nm AlSb, 5.9-nm In$_{0.25}$Ga$_{0.75}$Sb, InAs with thickness $t_{InAs}$, 50-nm AlSb, and 5-nm GaSb. The samples were processed into Hall bars with width of $W$ = 50 μm and voltage-probe distance of $L$ = 180 μm. Ohmic contacts were made after etching down to the InAs layer, depositing Ti/Au (10 nm/100 nm), and lift off, without annealing. A Ti/Au (20 nm/280 nm) front gate, fabricated on an atomic-layer deposited 40-nm-thick Al$_2$O$_3$ insulator, covers the active region of the Hall bar including the boundaries with the Ohmic contacts.

Fit parameters for Fig. 2(c)

Parameters used to fit the data in Fig. 2(c) are $c_F/e$ = 5.0 × 10$^{15}$ m$^{-2}$/V, $c_B/e$ = 7.5 × 10$^{14}$ m$^{-2}$/V, and ($\varepsilon_{InAs}$, $\varepsilon_{GaSb}$, $\varepsilon_{InSb}$) = (15.5, 15.7, 16.8)$\varepsilon_0$, where $\varepsilon_0$ is the vacuum permittivity. $c_M$ was calculated as $c_M = 1/(t_{InAs}/2\varepsilon_{InAs} + t_{InGaSb}/2\varepsilon_{InGaSb})$. The best fit was obtained for $m_e^*$ = 0.052$m_0$ with only weak dependence on hole effective mass for $m_h^*$ = (0.4 ± 0.3)$m_0$.